\documentclass[twocolumn,showpacs,preprintnumbers,amsmath,amssymb]{revtex4}
\usepackage{docs}

\usepackage{graphicx}
\usepackage{dcolumn}

\usepackage{bm}

\begin{document}
\title{Wormhole geometries in $f(R)$ modified
theories of gravity}

\author{Francisco S. N. Lobo}
\email{flobo@cii.fc.ul.pt} \affiliation{Centro de F\'{i}sica
Te\'{o}rica e Computacional, Faculdade de Ci\^{e}ncias da
Universidade de Lisboa, Avenida Professor Gama Pinto 2, P-1649-003
Lisboa, Portugal}

\author{Miguel A. Oliveira}
\email{miguel@cosmo.fis.fc.ul.pt}\affiliation{Centro de Astronomia
e Astrof\'{\i}sica da Universidade de Lisboa, Campo Grande, Ed. C8
1749-016 Lisboa, Portugal}

\date{\today}

\begin{abstract}

In this work, we construct traversable wormhole geometries in the
context of $f(R)$ modified theories of gravity. We impose that the
matter threading the wormhole satisfies the energy conditions, so
that it is the effective stress-energy tensor containing higher
order curvature derivatives that is responsible for the null
energy condition violation. Thus, the higher order curvature
terms, interpreted as a gravitational fluid, sustain these
non-standard wormhole geometries, fundamentally different from
their counterparts in general relativity. In particular, by
considering specific shape functions and several equations of
state, exact solutions for $f(R)$ are found.

\end{abstract}


\maketitle

\section{Introduction}

Various independent high-precision observational data have
confirmed with startling evidence that the Universe is undergoing
a phase of accelerated expansion~\cite{observations}. Several
candidates have been proposed in the literature to explain this
phenomenon, ranging from dark energy models to modified theories
of gravity. In the latter context, one may assume that at large
scales Einstein's theory of general relativity breaks down, and a
more general action describes the gravitational field. The
Einstein field equation of general relativity was first derived
from an action principle by Hilbert, by adopting a linear function
of the scalar curvature, $R$, in the gravitational Lagrangian
density. However, there are no a priori reasons to restrict the
gravitational Lagrangian to this form, and indeed several
generalizations have been proposed. In particular, a more general
modification of the Einstein-Hilbert gravitational Lagrangian
density involving an arbitrary function of the scalar invariant,
$f(R)$, was considered in \cite{Bu70}, and further developed in
\cite{Ke81}.

In this context, a renaissance of $f(R)$ modified theories of
gravity has been verified in an attempt to explain the late-time
accelerated expansion of the Universe (see Refs.
\cite{modgrav:review} for a review). Earlier interest in $f(R)$
theories was motivated by inflationary scenarios as for instance,
in the Starobinsky model, where $f(R)=R-\Lambda + \alpha R^2$ was
considered \cite{Starobinsky:1980te}. In fact, it was shown that
the late-time cosmic acceleration can be indeed explained within
the context of $f(R)$ gravity \cite{Carroll:2003wy}. Furthermore,
the conditions of viable cosmological models have been derived
\cite{viablemodels}, and an explicit coupling of an arbitrary
function of $R$ with the matter Lagrangian density has also been
explored \cite{coupling}. Relative to the Solar System regime,
severe weak field constraints seem to rule out most of the models
proposed so far \cite{solartests,Olmo07}, although viable models
do exist \cite{solartests2}. In the context of dark matter, the
possibility that the galactic dynamics of massive test particles
may be understood without the need for dark matter was also
considered in the framework of $f(R)$ gravity models
\cite{darkmatter}.

The metric formalism is usually considered in the literature,
which consists in varying the action with respect to $g^{\mu\nu}$.
However, other alternative approaches have been considered in the
literature, namely, the Palatini
formalism~\cite{Palatini,Sotiriou:2006qn}, where the metric and
the connections are treated as separate variables; and the
metric-affine formalism, where the matter part of the action now
depends and is varied with respect to the
connection~\cite{Sotiriou:2006qn}. The action for $f(R)$ modified
theories of gravity is given by
\begin{equation}
S=\frac{1}{2\kappa}\int d^4x\sqrt{-g}\;f(R)+S_M(g^{\mu\nu},\psi)
\,,
\end{equation}
where $\kappa =8\pi G$; throughout this work we consider
$\kappa=1$ for notational simplicity. $S_M(g^{\mu\nu},\psi)$ is
the matter action, defined as $S_M=\int d^4x\sqrt{-g}\;{\cal
L}_m(g_{\mu\nu},\psi)$, where ${\cal L}_m$ is the matter
Lagrangian density, in which matter is minimally coupled to the
metric $g_{\mu\nu}$ and $\psi$ collectively denotes the matter
fields.

Now, using the metric approach, by varying the action with respect
to $g^{\mu\nu}$, provides the following field equation
\begin{equation}
FR_{\mu\nu}-\frac{1}{2}f\,g_{\mu\nu}-\nabla_\mu \nabla_\nu
F+g_{\mu\nu}\Box F=\,T^m_{\mu\nu} \,,
    \label{field:eq}
\end{equation}
where $F\equiv df/dR$. Considering the contraction of Eq.
(\ref{field:eq}), provides the following relationship
\begin{equation}
FR-2f+3\,\Box F=\,T \,,
 \label{trace}
\end{equation}
which shows that the Ricci scalar is a fully dynamical degree of
freedom, and $T=T^{\mu}{}_{\mu}$ is the trace of the stress-energy
tensor.

In this work, we extend the analysis of static and spherically
symmetric spacetimes considered in the literature (for instance,
see \cite{SSSsolution}), and analyze traversable wormhole
geometries in $f(R)$ modified theories of gravity. Wormholes are
hypothetical tunnels in spacetime, possibly through which
observers may freely traverse. However, it is important to
emphasize that these solutions are primarily useful as
``gedanken-experiments'' and as a theoretician's probe of the
foundations of general relativity. In classical general
relativity, wormholes are supported by exotic matter, which
involves a stress-energy tensor that violates the null energy
condition (NEC) \cite{Morris:1988cz,Visser}. Note that the NEC is
given by $T_{\mu\nu}k^\mu k^\nu \geq 0$, where $k^\mu$ is {\it
any} null vector. Thus, it is an important and intriguing
challenge in wormhole physics to find a realistic matter source
that will support these exotic spacetimes. Several candidates have
been proposed in the literature, amongst which we refer to
solutions in higher dimensions, for instance in
Einstein-Gauss-Bonnet theory \cite{EGB1,EGB2}, wormholes on the
brane \cite{braneWH1}; solutions in Brans-Dicke theory
\cite{Nandi:1997en,Anchordoqui:1996jh,Agnese:1995kd}; wormhole
solutions in semi-classical gravity (see Ref.
\cite{Garattini:2007ff} and references therein); exact wormhole
solutions using a more systematic geometric approach were found
\cite{Boehmer:2007rm}; geometries supported by equations of state
responsible for the cosmic acceleration \cite{phantomWH},
solutions in conformal Weyl gravity were found \cite{Weylgrav},
and thin accretion disk observational signatures were also
explored \cite{Harko:2008vy}, etc (see Refs.
\cite{Lemos:2003jb,Lobo:2007zb} for more details and
\cite{Lobo:2007zb} for a recent review).

Thus, we explore the possibility that wormholes be supported by
$f(R)$ modified theories of gravity. It is an effective stress
energy, which may be interpreted as a gravitational fluid, that is
responsible for the null energy condition violation, thus
supporting these non-standard wormhole geometries, fundamentally
different from their counterparts in general relativity. We also
impose that the matter threading the wormhole satisfies the energy
conditions.

This paper is organized in the following manner: In Sec.
\ref{Sec:II}, the spacetime metric, the effective field equations
and the energy condition violations in the context of $f(R)$
modified theories of gravity are analyzed in detail. In Sec.
\ref{Sec:III}, specific solutions are explored, and we conclude in
Sec. \ref{Sec:conclusion}.

\section{Wormhole geometries in $f(R)$ gravity}\label{Sec:II}

\subsection{Spacetime metric and gravitational field equations}

Consider the wormhole geometry given by the following static and
spherically symmetric metric
\begin{equation}
ds^2=-e^{2\Phi(r)}dt^2+\frac{dr^2}{1-b(r)/r}+r^2\,(d\theta^2 +\sin
^2{\theta} \, d\phi ^2) \,,
    \label{metric}
\end{equation}
where $\Phi(r)$ and $b(r)$ are arbitrary functions of the radial
coordinate, $r$, denoted as the redshift function, and the shape
function, respectively \cite{Morris:1988cz}. The radial coordinate
$r$ is non-monotonic in that it decreases from infinity to a
minimum value $r_0$, representing the location of the throat of
the wormhole, where $b(r_0)=r_0$, and then it increases from $r_0$
back to infinity.

A fundamental property of a wormhole is that a flaring out
condition of the throat, given by $(b-b^{\prime}r)/b^{2}>0$, is
imposed \cite{Morris:1988cz}, and at the throat
$b(r_{0})=r=r_{0}$, the condition $b^{\prime}(r_{0})<1$ is imposed
to have wormhole solutions. It is precisely these restrictions
that impose the NEC violation in classical general relativity.
Another condition that needs to be satisfied is $1-b(r)/r>0$. For
the wormhole to be traversable, one must demand that there are no
horizons present, which are identified as the surfaces with
$e^{2\Phi}\rightarrow0$, so that $\Phi(r)$ must be finite
everywhere. In the analysis outlined below, we consider that the
redshift function is constant, $\Phi'=0$, which simplifies the
calculations considerably, and provide interesting exact wormhole
solutions (If $\Phi'\neq 0$, the field equations become forth
order differential equations, and become quite intractable).

The trace equation (\ref{trace}) can be used to simplify the field
equations and then can be kept as a constraint equation. Thus,
substituting the trace equation into Eq. (\ref{field:eq}), and
re-organizing the terms we end up with the following gravitational
field equation
\begin{equation}
G_{\mu\nu}\equiv R_{\mu\nu}-\frac{1}{2}R\,g_{\mu\nu}= T^{{\rm
eff}}_{\mu\nu} \,,
    \label{field:eq2}
\end{equation}
where the effective stress-energy tensor is given by $T^{{\rm
eff}}_{\mu\nu}= T^{(c)}_{\mu\nu}+\tilde{T}^{(m)}_{\mu\nu}$. The
term $\tilde{T}^{(m)}_{\mu\nu}$ is given by
\begin{equation}
\tilde{T}^{(m)}_{\mu\nu}=T^{(m)}_{\mu\nu}/F \,,
\end{equation}
and the curvature stress-energy tensor, $T^{(c)}_{\mu\nu}$, is
defined as
\begin{eqnarray}
T^{(c)}_{\mu\nu}=\frac{1}{F}\left[\nabla_\mu \nabla_\nu F
-\frac{1}{4}g_{\mu\nu}\left(RF+\Box F+T\right) \right]    \,.
    \label{gravfluid}
\end{eqnarray}

It is also interesting to consider the conservation law for the
above curvature stress-energy tensor. Taking into account the
Bianchi identities, $\nabla^\mu G_{\mu\nu}=0$, and the
diffeomorphism invariance of the matter part of the action, which
yields $\nabla^\mu T^{(m)}_{\mu\nu}=0$, we verify that the
effective Einstein field equation provides the following
conservation law
\begin{equation}\label{conserv-law}
\nabla^\mu T^{(c)}_{\mu\nu}=\frac{1}{F^2}
T^{(m)}_{\mu\nu}\nabla^\mu F  \,.
\end{equation}

Relative to the matter content of the wormhole, we impose that the
stress-energy tensor that threads the wormhole satisfies the
energy conditions, and is given by the following anisotropic
distribution of matter
\begin{equation}
T_{\mu\nu}=(\rho+p_t)U_\mu \, U_\nu+p_t\,
g_{\mu\nu}+(p_r-p_t)\chi_\mu \chi_\nu \,,
\end{equation}
where $U^\mu$ is the four-velocity, $\chi^\mu$ is the unit
spacelike vector in the radial direction, i.e.,
$\chi^\mu=\sqrt{1-b(r)/r}\,\delta^\mu{}_r$. $\rho(r)$ is the
energy density, $p_r(r)$ is the radial pressure measured in the
direction of $\chi^\mu$, and $p_t(r)$ is the transverse pressure
measured in the orthogonal direction to $\chi^\mu$. Taking into
account the above considerations, the stress-energy tensor is
given by the following profile: $T^{\mu}{}_{\nu}={\rm
diag}[-\rho(r),p_r(r),p_t(r),p_t(r)]$.

Thus, the effective field equation (\ref{field:eq2}) provides the
following relationships
\begin{eqnarray}
\frac{b'}{r^2}&=&\frac{\rho}{F}+\frac{H}{F}
  \,,    \label{fieldtt}
     \\
-\frac{b}{r^3}&=&\frac{p_r}{F}+\frac{1}{F}\Bigg\{\left(1-\frac{b}{r}\right)
\times
   \nonumber    \\
&&\times\left[F'' -F'\frac{b'r-b}{2r^2(1-b/r)}\right] -H\Bigg\}
  \,,  \label{fieldrr} \\
-\frac{b'r-b}{2r^3}
     &=&\frac{p_t}{F}+\frac{1}{F}\left[\left(1-\frac{b}{r}\right)
     \frac{F'}{r}
     -H\right]
     \label{fieldthetatheta}  \,,
\end{eqnarray}
where the prime denotes a derivative with respect to the radial
coordinate, $r$. The term $H=H(r)$ is defined as
\begin{equation}
H(r)=\frac{1}{4}\left(FR+\Box F +T\right) \,,
\end{equation}
for notational simplicity. The curvature scalar, $R$, is given by
\begin{eqnarray}
R&=& \frac{2b'}{r^2}
    \,,
    \label{Ricciscalar}
\end{eqnarray}
and $\Box F$ is provided by the following expression
\begin{equation}
\Box F=\left(1-\frac{b}{r}\right)\left[F''
-\frac{b'r-b}{2r^2(1-b/r)}\,F'+\frac{2F'}{r}\right] \,.
\end{equation}

Note that the gravitational field equations
(\ref{fieldtt})-(\ref{fieldthetatheta}), can be reorganized to
yield the following relationships:
\begin{eqnarray}
\label{generic1} \rho&=&\frac{Fb'}{r^2}\,,
       \\
\label{generic2}
p_r&=&-\frac{bF}{r^3}+\frac{F'}{2r^2}(b'r-b)-F''\left(1-\frac{b}{r}\right)
     \,,   \\
\label{generic3}
p_t&=&-\frac{F'}{r}\left(1-\frac{b}{r}\right)+\frac{F}{2r^3}(b-b'r)\,,
\end{eqnarray}
which are the generic expressions of the matter threading the
wormhole, as a function of the shape function and the specific
form of $F(r)$. Thus, by specifying the above functions, one
deduces the matter content of the wormhole.

One may now adopt several strategies to solve the field equations.
For instance, if $b(r)$ is specified, and using a specific
equation of state $p_r=p_r(\rho)$ or $p_t=p_t(\rho)$ one can
obtain $F(r)$ from the gravitational field equations and the
curvature scalar in a parametric form, $R(r)$, from its definition
via the metric. Then, once $T=T^{\mu}{}_{\mu}$ is known as a
function of $r$, one may in principle obtain $f(R)$ as a function
of $R$ from Eq. (\ref{trace}).

\subsection{Energy condition violations}

A fundamental point in wormhole physics is the energy condition
violations, as mentioned above. However, a subtle issue needs to
be pointed out in modified theories of gravity, where the
gravitational field equations differ from the classical
relativistic Einstein equations. More specifically, we emphasize
that the energy conditions arise when one refers back to the
Raychaudhuri equation for the expansion where a term
$R_{\mu\nu}k^\mu k^\nu$ appears, with $k^\mu$ any null vector. The
positivity of this quantity ensures that geodesic congruences
focus within a finite value of the parameter labelling points on
the geodesics. However, in general relativity, through the
Einstein field equation one can write the above condition in terms
of the stress-energy tensor given by $T_{\mu\nu}k^\mu k^\nu \ge
0$. In any other theory of gravity, one would require to know how
one can replace $R_{\mu\nu}$ using the corresponding field
equations and hence using matter stresses. In particular, in a
theory where we still have an Einstein-Hilbert term, the task of
evaluating $R_{\mu\nu}k^\mu k^\nu$ is trivial. However, in $f(R)$
modified theories of gravity under consideration, things are not
so straightforward.

Now the positivity condition, $R_{\mu\nu}k^\mu k^\nu \geq 0$, in
the Raychaudhuri equation provides the following form for the null
energy condition $T^{{\rm eff}}_{\mu\nu} k^\mu k^\nu\geq 0$,
through the modified gravitational field equation
(\ref{field:eq2}), and it this relationship that will be used
throughout this work. For this case, in principle, one may impose
that the matter stress-energy tensor satisfies the energy
conditions and the respective violations arise from the higher
derivative curvature terms $T^{(c)}_{\mu\nu}$. Another approach to
the energy conditions considers in taking the condition
$T_{\mu\nu} k^\mu k^\nu\ge 0$ at face value. Note that this is
useful as using local Lorentz transformations it is possible to
show that the above condition implies that the energy density is
positive in all local frames of reference. However, if the theory
of gravity is chosen to be non-Einsteinian, then the assumption of
the above condition does not necessarily imply focusing of
geodesics. The focusing criterion is different and will follow
from the nature of $R_{\mu\nu} k^\mu k^\nu$.

Thus, considering a radial null vector, the violation of the NEC,
i.e., $T_{\mu\nu}^{{\rm eff}}\,k^\mu k^\nu < 0$ takes the
following form
\begin{equation}
\rho^{{\rm eff}}+p_r^{{\rm
eff}}=\frac{\rho+p_r}{F}+\frac{1}{F}\left(1-\frac{b}{r}\right)
\left[F''-F'\frac{b'r-b}{2r^2(1-b/r)}\right],
     \label{NECeff}
\end{equation}
where $\rho^{{\rm eff}}+p_r^{{\rm eff}}<0$. Using the
gravitational field equations, inequality (\ref{NECeff}) takes the
familiar form
\begin{equation}
\rho^{{\rm eff}}+p_r^{{\rm eff}}=\frac{b'r-b}{r^3}\,,
     \label{NECeff2}
\end{equation}
which is negative by taking into account the flaring out
condition, i.e., $(b'r-b)/b^2<0$, considered above.

At the throat, one has the following relationship
\begin{equation}
\rho^{{\rm eff}}+p_r^{{\rm eff}}|_{r_0}=
\frac{\rho+p_r}{F}\Big|_{r_0}+
\frac{1-b'(r_0)}{2r_0}\frac{F'}{F}\Big|_{r_0}<0 \,.
     \label{NECeffb}
\end{equation}
It is now possible to find the following generic relationships for
$F$ and $F'$ at the throat: $F'_0<-2r_0(\rho+p_r)|_{r_0}/(1-b')$
if $F>0$; and $F'_0>-2r_0(\rho+p_r)|_{r_0}/(1-b')$ if $F<0$.

Consider that the matter threading the wormhole obeys the energy
conditions. To this effect, imposing the weak energy condition
(WEC), given by $\rho \geq 0$ and $\rho + p_r \geq 0$, then Eqs.
(\ref{generic1})-(\ref{generic2}) yield the following
inequalities:
\begin{eqnarray}
\frac{Fb'}{r^2}\geq 0 \,, \label{WEC1}
  \\
\frac{(2F+rF')(b'r-b)}{2r^2}-F''\left(1-\frac{b}{r}\right)\geq 0
\,, \label{WEC2}
\end{eqnarray}
respectively.

Thus, if one imposes that the matter threading the wormhole
satisfies the energy conditions, we emphasize that it is the
higher derivative curvature terms that sustain the wormhole
geometries. Thus, in finding wormhole solutions it is fundamental
that the functions $f(R)$ obey inequalities (\ref{NECeff}) and
(\ref{WEC1})-(\ref{WEC2}).

\section{Specific solutions}\label{Sec:III}

In this section, we are mainly interested in adopting the strategy
of specifying the shape function $b(r)$, which yields the
curvature scalar in a parametric form, $R(r)$, from its definition
via the metric, given by Eq. (\ref{Ricciscalar}). Then, using a
specific equation of state $p_r=p_r(\rho)$ or $p_t=p_t(\rho)$, one
may in principle obtain $F(r)$ from the gravitational field
equations. Finally, once $T=T^{\mu}{}_{\mu}$ is known as a
function of $r$, one may in principle obtain $f(R)$ as a function
of $R$ from Eq. (\ref{trace}).

\subsection{Traceless stress-energy tensor}

An interesting equation of state is that of the traceless
stress-energy tensor, which is usually associated to the Casimir
effect, with a massless field. Note that the Casimir effect is
sometimes theoretically invoked to provide exotic matter to the
system considered at hand. Thus, taking into account the traceless
stress-energy tensor, $T=-\rho+p_r+2p_t=0$, provides the following
differential equation
\begin{equation}
F''\left(1-\frac{b}{r}\right)-\frac{b'r+b-2r}{2r^2}F'-
\frac{b'r-b}{2r^3}F=0\,. \label{diffeqT}
\end{equation}
In principle, one may deduce $F(r)$ by imposing a specific shape
function, and inverting Eq. (\ref{Ricciscalar}), i.e., $R(r)$, to
find $r(R)$, the specific form of $f(R)$ may be found from the
trace equation (\ref{trace}).

For instance, consider the specific shape function given by
$b(r)=r_0^2/r$. Thus, Eq. (\ref{diffeqT}) provides the following
solution
\begin{eqnarray}
F(r)&=&C_1
\sinh\left[\sqrt{2}\,\arctan\left(\frac{r_0}{\sqrt{r^2-r_0^2}}
\right)\right]
   \nonumber   \\
&&+C_2\cosh\left[\sqrt{2}\,\arctan\left(\frac{r_0}{\sqrt{r^2-r_0^2}}
\right)\right]\,.
\end{eqnarray}

The stress-energy tensor profile threading the wormhole is given
by the following relationships
\begin{widetext}
\begin{eqnarray}
\rho(r)&=&- \frac{r_0^2}{r^4}\Bigg\{C_1
\sinh\left[\sqrt{2}\,\arctan\left(\frac{r_0}{\sqrt{r^2-r_0^2}}
\right)\right]
   +C_2\cosh\left[\sqrt{2}\,\arctan\left(\frac{r_0}{\sqrt{r^2-r_0^2}}
\right)\right]\Bigg\}\,,\\
p_r(r)&=&- \frac{r_0}{r^4}\Bigg\{\left(2C_2\sqrt{2(r^2-r_0^2)}+
3r_0C_1\right)
\sinh\left[\sqrt{2}\,\arctan\left(\frac{r_0}{\sqrt{r^2-r_0^2}}
\right)\right]
 \nonumber   \\
&&  +\left(2C_1\sqrt{2(r^2-r_0^2)}+
3r_0C_2\right)\cosh\left[\sqrt{2}\,\arctan\left(\frac{r_0}{\sqrt{r^2-r_0^2}}
\right)\right]\Bigg\}\,,\\
p_t(r)&=& \frac{r_0}{r^4}\Bigg\{\left(C_2\sqrt{2(r^2-r_0^2)}+
r_0C_1\right)
\sinh\left[\sqrt{2}\,\arctan\left(\frac{r_0}{\sqrt{r^2-r_0^2}}
\right)\right]
 \nonumber   \\
&&  +\left(C_1\sqrt{2(r^2-r_0^2)}+
r_0C_2\right)\cosh\left[\sqrt{2}\,\arctan\left(\frac{r_0}{\sqrt{r^2-r_0^2}}
\right)\right]\Bigg\}\,.
\end{eqnarray}
\end{widetext}

\begin{figure}[h]
\centering
  \includegraphics[width=2.8in]{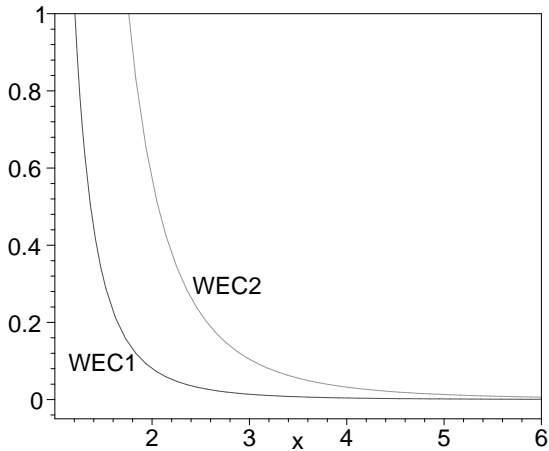}
  \caption{The stress-energy tensor satisfying the WEC, for the
  specific case of the traceless stress-energy tensor equation of
  state, and for the values $C_1=0$ and $C_2=-1$.
  We have considered the dimensionless quantities
  ${\rm WEC}1=r_0^2\rho$, ${\rm WEC}2=r_0^2(\rho+p_r)$ and $x=r/r_0$.}
 \label{Fig:WECT=0}
\end{figure}

One may now impose that the above stress-energy tensor satisfies
the WEC, which is depicted in Fig. \ref{Fig:WECT=0}, by
considering the values $C_1=0$ and $C_2=-1$.

\begin{figure}[h]
\centering
  \includegraphics[width=3.1in]{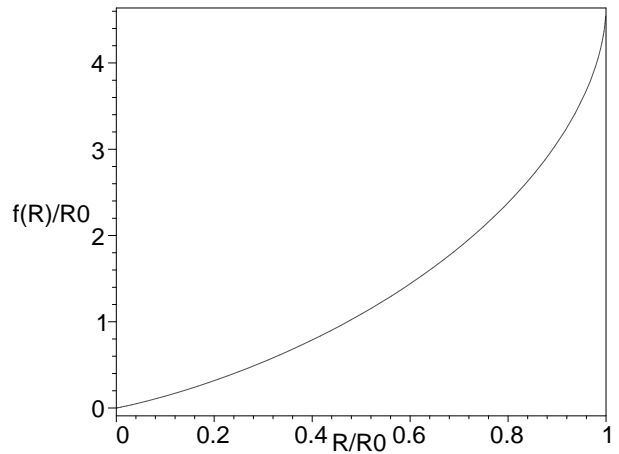}
  \caption{The specific form of $f(R)$, for the specific case of
  the traceless stress-energy tensor equation of state, by imposing
  the values $C_1=0$ and $C_2=-1$. The range is given by
  $0\leq R/R_0\leq 1$.}
 \label{Fig:WECT=0fR}
\end{figure}

For the specific shape function considered above, the Ricci
scalar, Eq. (\ref{Ricciscalar}), provides $R=-2r_0^2/r^4$ and is
now readily inverted to give $r=(-2r_0^2/R)^{1/4}$. It is also
convenient to define the Ricci scalar at the throat, and its
inverse provides $r_0=(-2/R_0)^{1/2}$. Substituting these
relationships into the consistency equation (\ref{trace}), the
specific form $f(R)$ is given by
\begin{eqnarray}
f(R)&=&-R\Bigg\{C_1\sinh\left[\sqrt{2}\arctan\left( \frac{1}
{\sqrt{\left(\frac{R_0}{R}\right)^{1/2}-1}} \right)\right]
    \nonumber   \\
&&\hspace{-0.5cm}+C_2\cosh\left[\sqrt{2}\arctan\left( \frac{1}
{\sqrt{\left(\frac{R_0}{R}\right)^{1/2}-1}}
\right)\right]\Bigg\}\,,
\end{eqnarray}
which is depicted in the Fig. \ref{Fig:WECT=0fR}, by imposing the
values $C_1=0$ and $C_2=-1$.

\subsection{Specific equation of state: $p_t=\alpha \rho$}

Many of the equations of state considered in the literature
involving the radial pressure and the energy density, such as the
linear equation of state $p_r=\alpha \rho$, provide very complex
differential equations, so that it is extremely difficult to find
exact solutions. This is due to the presence of the term $F''$ in
$p_r$. Indeed, even considering isotropic pressures does not
provide an exact solution. Now, things are simplified if one
considers an equation of state relating the tangential pressure
and the energy density, so that the radial pressure is determined
through Eq. (\ref{generic2}). For instance, consider the equation
of state $p_t=\alpha \rho$, which provides the following
differential equation:
\begin{equation}
F'\left(1-\frac{b}{r}\right)-\frac{F}{2r^2}
\left[b-b'r(1+2\alpha)\right]=0\,.
\label{diffeq1}
\end{equation}
In principle, as mentioned above one may deduce $F(r)$ by imposing
a specific shape function, and inverting Eq. (\ref{Ricciscalar}),
i.e., $R(r)$, to find $r(R)$, the specific form of $f(R)$ may be
found from the trace equation (\ref{trace}). In the following
analysis we consider several interesting shape functions usually
applied in the literature.

\subsubsection{1. Specific shape function:
$b(r)=r_0^2/r$}

First, we consider the case of $b(r)=r_0^2/r$, so that Eq.
(\ref{diffeq1}) yields the following solution
\begin{equation}
F(r)=C_1\left(1-\frac{r_0^2}{r^2}\right)^{\frac{1}{2}+\frac{\alpha}{2}}
\,.
\end{equation}

\begin{figure}[h]
\centering
  \includegraphics[width=2.8in]{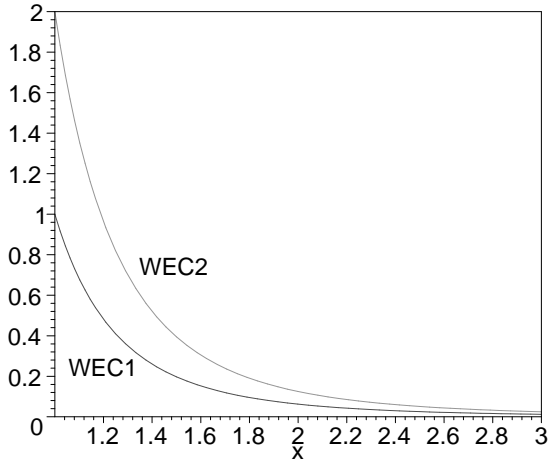}
  \caption{The stress-energy tensor satisfies the WEC,
  for the specific case of the equation of state $p_t=\alpha \rho$
  and considering the form function $b(r)=r_0^2/r$. We have
  imposed the values $C_1=-1$ and $\alpha=-1$, and
  considered the dimensionless quantities
  ${\rm WEC}1=r_0^2\rho$, ${\rm WEC}2=r_0^2(\rho+p_r)$ and $x=r/r_0$.}
 \label{Fig:WECptA}
\end{figure}

The gravitational field equations,
(\ref{generic1})-(\ref{generic3}), provide the stress-energy
tensor threading the wormhole, given by the following
relationships
\begin{eqnarray}
p_r(r)&=&\frac{C_1r_0^2}{r^6}\left(1-\frac{r_0^2}{r^2}\right)
^{-\frac{1}{2}+\frac{\alpha}{2}} \times
    \nonumber   \\
&&\times\left[2(r^2-r_0^2)+3\alpha
r^2-4r_0^2\alpha-r_0^2\alpha^2\right]\,,
   \\
p_t(r)&=&\alpha\rho(r)=-\frac{C_1r_0^2\alpha}{r^4}
\left(1-\frac{r_0^2}{r^2}\right)
^{\frac{1}{2}+\frac{\alpha}{2}} \,.
\end{eqnarray}
One may now impose that the above stress-energy tensor satisfies
the WEC, which is depicted in Fig. \ref{Fig:WECptA}, by imposing
the values $C_1=-1$ and $\alpha=-1$.

\begin{figure}[t]
\centering
  \includegraphics[width=3.1in]{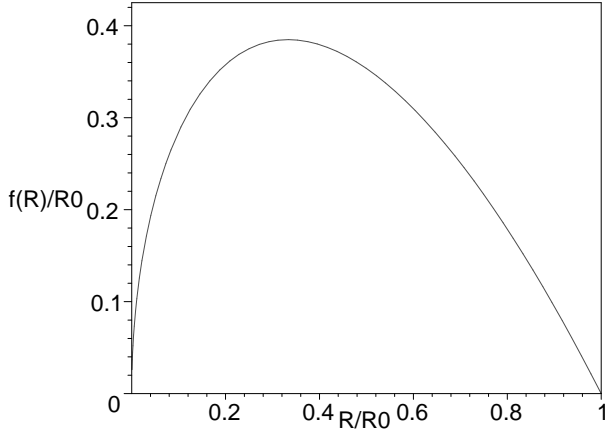}
  \caption{The profile of $f(R)$ is depicted for the
  specific case of the equation of state $p_t=\alpha \rho$
  and considering the form function $b(r)=r_0^2/r$. The values
  $C_1=-1$ and $\alpha=-1$ have been imposed,
  with the range given by $0\leq R/R_0\leq 1$.}
 \label{Fig:ptAfR}
\end{figure}

As in the previous case of the traceless stress-energy tensor, the
Ricci scalar, Eq. (\ref{Ricciscalar}), is given by $R=-2r_0^2/r^4$
and its inverse provides $r=(-2r_0^2/R)^{1/4}$. The inverse of the
Ricci scalar evaluated at the throat inverse is given by
$r_0=(-2/R_0)^{1/2}$. Substituting these relationships into the
consistency relationship (\ref{trace}), provides the specific form
of $f(R)$, which is given by
\begin{eqnarray}
f(R)&=&C_1R\left(1
-\sqrt{\frac{R}{R_0}}\right)^{\frac{\alpha}{2}-\frac{1}{2}} \times
   \nonumber   \\
&&\hspace{-1cm}\times\left[\sqrt{\frac{R}{R_0}}
(\alpha^2+2\alpha+2)+(\alpha+2) \right]
 \,.
\end{eqnarray}
This function is depicted in Fig. \ref{Fig:ptAfR} as $f(R)/R_0$ as
a function as $R/R_0$, for the values $C_1=-1$ and $\alpha=-1$.

\subsubsection{2. Specific shape function: $b=\sqrt{r_0r}$}

Consider now the case of $b=\sqrt{r_0r}$, so that Eq.
(\ref{diffeq1}) yields the following solution
\begin{equation}
F(r)=C_1\left(1-\sqrt{\frac{r_0}{r}}\,\right)^{\frac{1}{2}-\alpha}
\,.
\end{equation}

The stress-energy tensor profile threading the wormhole is given
by the following relationships
\begin{eqnarray}
    p_t(r)&=&\alpha\rho(r)=\frac{C_1 \alpha}{2r^2}
    \frac{\left(1-\sqrt{\frac{r_0}{r}}\right)^{\frac{1}{2}
    -\alpha}}{\sqrt{\frac{r_0}{r}}}\,, \\
    p_r(r)&=&-\frac{C_1r_0}{16r^3}
    \left(1-\sqrt{\frac{r_0}{r}}\right)^{-\frac{3}{2}-\alpha}\times
    \nonumber\\
    &&\hspace{-1.75cm}\times\bigg[10\sqrt{\frac{r}{r_0}}
    +\sqrt{\frac{r_0}{r}}
    \left(14\alpha+10\right)+
    \left(4\alpha^2-26\alpha+5\right)\bigg] \,.
\end{eqnarray}
Rather than consider plots of the WEC as before, we note that it
is possible to impose various specific values of $C_1$ and
$\alpha$ that do indeed satisfy the WEC.

Following the recipe prescribed above, the Ricci scalar is given
by $R=\sqrt{r_0}/r^{5/2}$ and is readily inverted to provide
$r=(\sqrt{r_0}/R)^{2/5}$. The inverse of the Ricci scalar at the
throat provides $r_0=1/\sqrt{R_0}$. Substituting these
relationships into the consistency relationship (\ref{trace}), the
specific form $f(R)$ is finally given by
\begin{widetext}
\begin{eqnarray}
f(R)&=&-\frac{1}{8}\frac{C_1}{ R^\frac{2}{5}
-2\left(RR_0\right)^\frac{1}{5}+R_0^\frac{2}{5}} \Bigg\{
\left(R_0^\frac{1}{5}-R^\frac{1}{5} \right)^{\frac{1}{2}-\alpha}
R^\frac{3-2\alpha}{10} R_0^\frac{-21+10\alpha}{40}\times
\nonumber\\
&&\times \left[-8R R_0^\frac{2}{5}+\right(11+10\alpha
\left)R^\frac{4}{5}R_0^\frac{3}{5}+ \right(2-22\alpha
+4\alpha^2\left)R^\frac{3}{5}R_0^\frac{4}{5}
+\left(-5+12\alpha-4\alpha^2\right)R^\frac{2}{5}R_0\right]\Bigg\}\,.
\end{eqnarray}
\end{widetext}

\subsubsection{3. Specific shape function:
$b(r)=r_0+\gamma^2r_0(1-r_0/r)$}

Finally, it is also of interest to consider the specific shape
function given by $b(r)=r_0+\gamma^2r_0(1-r_0/r)$, with
$0<\gamma<1$, so that Eq. (\ref{diffeq1}) provides the following
solution
\begin{eqnarray}
F(r)&=&C_1\left(r-\gamma^2r_0\right)^{\frac{1}{2}
\frac{\gamma^2-2\alpha-1}{\gamma^2-1}}
\nonumber \\
&&r^{-(\alpha+1)}\left(r-r_0\right)^{\frac{1}{2}
\frac{\gamma^2(1+2\alpha)-1}{\gamma^2-1}}
\end{eqnarray}
It is useful to write the last equation in the form
$F(r)=C_1X^ur^{-(\alpha+1)}Y^v$, where $X,\,Y,\,u,\, v$ are
defined as
\begin{eqnarray*}
&&X=r-\gamma^2r_0 \,, \qquad  Y=r-r_0\,,\\
&&u=\frac{\gamma^2-2\alpha-1}{2(\gamma^2-1)} \,, \qquad
v=\frac{\gamma^2(1+2\alpha)-1}{2(\gamma^2-1)}  \,.
\end{eqnarray*}

Thus, the stress energy tensor profile threading the wormhole is
given by the following expressions:
\begin{widetext}
\begin{eqnarray}\nonumber
  p_r(r)&=& \frac{C_1}{2r^3}\Bigg\{X^uY^v\bigg[r^{-\alpha}
  \left(2\alpha^2+6\alpha+4\right)+r^{-(1+\alpha)}r_0
  \left(-7\alpha+2\alpha^2\gamma^2-3\gamma^2-7\alpha\gamma^2-2\alpha^2
 -3\right)
 \\ \nonumber
 &&+r^{-(2+\alpha)}r_0^2\gamma^2\left(10\alpha^2+4\right)\bigg]
 +X^uY^{v-1}
 \bigg[r^{-\alpha}r_0v\left(-\gamma^2(5+4\alpha)-\alpha-5\right)
 +4r^{1-\alpha}v\left(1+\alpha\right)
 \\
 \nonumber
 &&+r^{-(1+\alpha)}r_0^2\gamma^2v\left(4\alpha+6\right)\bigg]
 +X^uY^{v-2}\bigg[2r^{-\alpha}r_0\gamma^2v\left(v-\alpha \right)
 +2r^{1-\alpha}r_0v\left(-v+\gamma^2+1\right)
 \\
 &&+2r^{2-\alpha}v\left(v+1\right)\bigg]+X^{u-1}Y^v
 \bigg[r^{-\alpha}r_0u\left(4\alpha+5\right)\left(\gamma^2+1\right)
 -4r^{1-\alpha}\left(u-\alpha\right)\\\nonumber
 &&-r^{-(1+\alpha)}r_0^2\gamma^2u\left(4\alpha-6\right)\bigg]
 +X^{u-2}Y^v\big[2r^{-\alpha}r_0^2\gamma^2u\left(u-1\right)
 +2r^{1-\alpha}r_0u\left(1-u\right)
 +2r^{2-\alpha}u\left(u-1\right)\bigg]\\
 \nonumber
 &&+X^{u-1}Y^{v-1}\bigg[-4r^{-\alpha}r_0^2\gamma^2uv
 +4r^{1-\alpha}r_0uv\left(\gamma^2+1\right)
 -4r^{2-\alpha}uv\bigg]\Bigg\}\,,\\
  p_t(r)&=&\alpha\rho=C_1\alpha
  \gamma^2r_0^2X^ur^{-(5+\alpha)}Y^v\,.
\end{eqnarray}
\end{widetext}
As in the previous example, we will not depict the plot of the
functions, but simply note in passing that one may impose specific
values for the constants $\alpha$ and $C_1$ in order to satisfy
the WEC.

The Ricci scalar, Eq. (\ref{Ricciscalar}), provides
$R=2\gamma^2r_0^2/r^4$ and is now readily inverted to give
$r=(2\gamma^2r_0^2/R)^{1/4}$. The Ricci scalar at the throat is
given by $R_0=2\gamma^2/r_0^2$, and its inverse provides
$r_0=\gamma\sqrt{2/R_0}$. Substituting these relationships into
the consistency relationship (\ref{trace}), the specific form
$f(R)$ is given by
\begin{widetext}
\begin{eqnarray}
f(R)&=&\frac{C_1R}{2}\frac{\left(R_0
R\right)^{\frac{(\alpha+1)}{4}}}{
\gamma^2-\left(\frac{R_0}{R}\right)^\frac{1}{4}
(\gamma^2+1)+\left(\frac{R_0}{R}\right)^{\frac{1}{2}}}
\left[\frac{\left(\frac{R_0}{R}\right)^\frac{1}{4}
-\gamma^2}{R_0^\frac{1}{2}}\right]^{\frac{1}{2}\frac{\gamma^2
-2\alpha-1}{\gamma^2-1}}\left[\frac{\left(\frac{R_0}{R}
\right)^\frac{1}{4}-1}{R_0^\frac{1}{2}}\right]^{\frac{1}{2}
\frac{\gamma^2(1-2\alpha)-1}{\gamma^2-1}}\times
\nonumber\\
&&\left[2\gamma^2(\alpha^2+2\alpha+2)
-\left(\frac{R_0}{R}\right)^\frac{1}{4}(3\alpha+4)
(\gamma^2+1)+\left(\frac{R_0}{R}\right)^\frac{1}{2}
(2\alpha+4)\right]\,.
\end{eqnarray}
\end{widetext}

\section{Summary and Discussion}\label{Sec:conclusion}

In general relativity, the NEC violation is a fundamental
ingredient of static traversable wormholes. Despite this fact, it
was shown that for time-dependent wormhole solutions the null
energy condition and the weak energy condition can be avoided in
certain regions and for specific intervals of time at the throat
\cite{dynamicWH}. Nevertheless, in certain alternative theories to
general relativity, taking into account the modified Einstein
field equation, one may impose in principle that the stress energy
tensor threading the wormhole satisfies the NEC. However, the
latter is necessarily violated by an effective total stress energy
tensor. This is the case, for instance, in braneworld wormhole
solutions, where the matter confined on the brane satisfies the
energy conditions, and it is the local high-energy bulk effects
and nonlocal corrections from the Weyl curvature in the bulk that
induce a NEC violating signature on the brane \cite{braneWH1}.
Another particularly interesting example is in the context of the
$D$-dimensional Einstein-Gauss-Bonnet theory of gravitation
\cite{EGB1}, where it was shown that the weak energy condition can
be satisfied depending on the parameters of the theory.

In this work, we have explored the possibility that wormholes be
supported by $f(R)$ modified theories of gravity. We imposed that
the matter threading the wormhole satisfies the energy conditions,
and it is the higher order curvature derivative terms, that may be
interpreted as a gravitational fluid, that support these
non-standard wormhole geometries, fundamentally different from
their counterparts in general relativity. In the analysis outlined
above, we considered a constant redshift function, which
simplified the calculations considerably, yet provides interesting
enough exact solutions. One may also generalize the results of
this paper by considering $\Phi'\neq 0$, although the field
equations become forth order differential equations, and become
quite intractable. The strategy adopted to solve the field
equations was essentially to specify $b(r)$, and considering
specific equation of state, the function $F(r)$ was deduced from
the gravitational field equations, while the curvature scalar in a
parametric form, $R(r)$, was obtained from its definition via the
metric. Then, deducing $T=T^{\mu}{}_{\mu}$ as a function of $r$,
exact solutions of $f(R)$ as a function of $R$ from the trace
equation were found.

Furthermore, we note that $f(R)$ modified theories of gravity are
equivalent to a Brans-Dicke theory with a coupling parameter
$w=0$, and a specific potential related to the function $f(R)$ and
its derivative. In this context, it was shown that static wormhole
solutions in the vacuum Brans-Dicke theory only exist in a narrow
interval of the coupling parameter \cite{Nandi:1997en}, namely,
$-3/2<w<-4/3$. However, we point out that this result is only
valid for vacuum solutions and for a specific choice of an
integration constant of the field equations given by
$C(w)=-1/(w+2)$. The latter relationship was derived on the basis
of a post-Newtonian weak field approximation, and it is important
to emphasize that there is no reason for it to hold in the
presence of compact objects with strong gravitational fields.

Another issue that needs to be mentioned, is that the
above-mentioned interval imposed on $w$ was obtained by
considering the violation of the WEC (recall that the WEC imposes
$\rho \geq 0$ and $\rho+p_r \geq 0$). Now the authors in
\cite{Nandi:1997en} obtained the respective constraints on $w$ by
considering negative energy densities, i.e., $\rho < 0$. This is
not a necessary condition, as one may consider positive energy
densities and in alternative impose the condition $\rho+p_r < 0$,
which violates the WEC, and consequently the NEC. Note that this
is justified as the fundamental ingredient in wormhole physics is
the violation of the NEC, and not the imposition of negative
energy densities. In principle, this condition combined with an
adequate choice of $C(w)$ could provide a different viability and
less restrictive interval (including the value $w=0$) from the
case of $-3/2<w<-4/3$ considered in \cite{Nandi:1997en}.

For the vacuum case considered in the present paper, we note that
there are no viable solutions, as now we have three gravitational
field equations and two arbitrary functions, $b(r)$ and $F(r)$, so
that the system is over-determined. This difficulty can be
surpassed by considering the general case of $\Phi'(r)\neq 0$, but
now it is impossible to find an exact analytical solution, and
numerical methods are needed to solve the system of equations.
However, in the presence of matter things are totally different,
as this adds additional degrees of freedom. Thus, in principle one
may construct a whole plethora of wormhole solutions (a specific
equation of state was considered in \cite{Anchordoqui:1996jh}), in
addition to adequately choosing $C(w)$ in Brans-Dicke theory. Work
along these lines in presently underway.

\section*{Acknowledgements}
MO acknowledges financial support from a grant attributed by
Centro de Astronomia e Astrof\'{\i}sica da Universidade de Lisboa
(CAAUL), and financed by Funda\c{c}\~ao para a Ci\^{e}ncia e
Tecnologia (FCT).



\end{document}